\documentclass[twocolumn,prl]{revtex4}

  \usepackage[dvips]{graphicx}
\usepackage{dcolumn}
\usepackage{amsmath}
\usepackage{latexsym}

\begin{document}

\title[Short Title]{Single flux quantum circuits
with damping based on dissipative transmission lines}

\author{E.\,M.~Tolkacheva}
 \altaffiliation[Present address: ]{D-wave Systems Inc., 100-4401
 Still Creek Drive, Burnaby, B.C. V5C 6G9, Canada}
\author{D.\,V.~Balashov}
\author{M.\,I.~Khabipov}
\author{A.\,B.~Zorin}

\affiliation{Physikalisch-Technische Bundesanstalt, Bundesallee
100, 38116 Braunschweig, Germany}%

\date{April 2, 2008}

\begin{abstract}
We propose and demonstrate the functioning of a special Rapid Single
Flux Quantum (RSFQ) circuit with frequency-dependent damping. This
damping is achieved by shunting individual Josephson junctions by
pieces of open-ended $RC$ transmission lines. Our circuit includes a
toggle flip-flop cell, Josephson transmission lines transferring
single flux quantum pulses to and from this cell, as well as DC/SFQ
and SFQ/DC converters. Due to the desired frequency-dispersion in
the $RC$ line shunts which ensures sufficiently weak damping at low
frequencies, such circuits are well-suited for integrating with the
flux/phase Josephson qubit and enable its efficient control.


\end{abstract}
\maketitle

\section{Introduction}

The Josephson circuits operating on Single Flux Quanta (SFQ), the
so-called Rapid Single Flux Quantum (RSFQ) circuits
\cite{LikharevSemenov} of special design, are intensively
investigated as they are the most promising on-chip system enabling
the control and readout of the superconducting qubits
\cite{SemenovAverin, WendinShumeiko}. This choice is primarily based
on the well developed fabrication technology of these circuits,
allowing them to be implemented in the same process together with
the Josephson qubits. Moreover, the capability of the RSFQ circuits
to function at the same low temperature as the qubits is another
advantage of this approach. Among these RSFQ circuits are the
balanced comparators \cite{Wulf, Savin, Ohki}, SQUIDs \cite{Li}, the
ballistic Josephson Transmission Lines (JTL) \cite{Averin1, Fedorov}
enabling quantum readout of the qubit, and the RSFQ circuits
generating signals controlling the qubit state
\cite{Crankshaw,Castellano}.

\begin{figure}[t]
\begin{center}
\includegraphics[width=7.5cm]{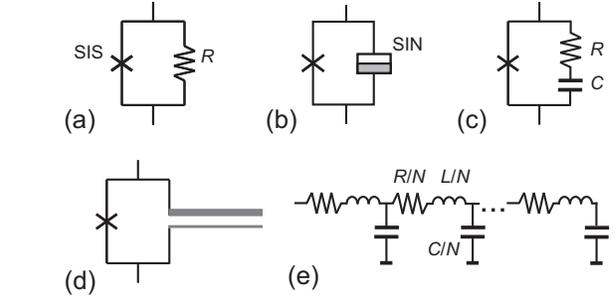}
\caption{Different ways of realizing the damping in Josephson SIS
junctions: (a) resistive shunting, (b) shunting by an SIN tunnel
element, (c) shunting by an $RC$ circuit, and (d) shunting by an
open-ended dissipative transmission line. (e) The equivalent
electric diagram of such transmission line modeled by the ladder
circuit consisting of sufficiently large number $N$ of elementary
$RLC$ sections.} \label{fig:all-shunts}
\end{center}
\end{figure}

One of the most important issues in integrating the RSFQ circuits
with a qubit is their strong back action on the qubit. On the one
hand, the operating principle of the RSFQ circuits assumes the
availability of significant damping which is usually ensured by the
low-ohmic resistive shunting $R$ of the Josephson Superconductor -
Insulator - Superconductor (SIS) tunnel junctions (see the electric
diagram in Fig.\,\ref{fig:all-shunts}(a), where the cross denotes
the junction itself with a presumably very high quasiparticle
resistance $R_{\textrm{qp}}\gg R$). On the other hand, such
resistive shunts generate significant current noise with the white
spectrum even in the quiescent (zero-voltage) state of the Josephson
junctions. This noise is admixed to the bias and output control
signals and, therefore, may dramatically decohere the qubit
\cite{Makhlin,Wal}. A possible solution to this problem is the
application of frequency-dependent elements which ensure sufficient
damping at plasma resonance frequency $\nu_p = \omega_p/2\pi \sim
40$\,GHz of the junctions of the RSFQ circuit, but have a relatively
low damping and, therefore, a low noise at frequencies below the
characteristic frequency of the qubit $\nu_{01} = (E_1-E_0)/h
\lesssim 10$\,GHz. As we had shown earlier \cite{Z-APL, Z-PRB}, to
realize such damping, one can use non-linear shunts on the basis of
Superconductor - Insulator - Normal metal (SIN) tunnel junctions
(see Fig.\,\ref{fig:all-shunts}(b)). Although these shunts may
ensure both a remarkably large high-frequency damping and a
significant suppression of the low-frequency noise due to a
presumably small sub-gap leakage current in the SIN junctions
\cite{Lotkhov-PhysC}, the fabrication technology of these hybrid
SIS+SIN circuits is rather difficult.

Recently,  a simpler realization of RSFQ networks with a
frequency-dependent damping based on shunting the junctions by serial $RC$
circuits (see Fig.\,\ref{fig:all-shunts}(c)) has been proposed in
Ref. \cite{VTT} and verified in Ref. \cite{VTT-2}. This linear
circuit consisting of thin-film resistor and
plate on-chip capacitor is relatively simple for fabrication.
Application of such shunting elements was beneficial in the
low-noise dc SQUIDs used in magnetometry \cite{VTT-3} and in the
readout SQUIDs for the Josephson qubit \cite{Robertson}.

In this paper, we propose to damp an RSFQ circuit by shunting its
junctions by pieces of dissipative transmission lines (see
Fig.\,\ref{fig:all-shunts}(d)-(e)). We report the results of
numerical simulations and of the measurements of a Nb RSFQ circuit
provided with such damping. Our experimental circuit includes Toggle Flip-Flop
(TFF), Josephson transmission lines (JTL) and DC/SFQ and SFQ/DC
converters. We will show that a microstrip line shunts made of a
resistive film which is normally used for the fabrication of
shunting resistors can ensure a stable operation of the circuit with
a desirable frequency-dependent damping and give advantages in the
suppression of noise at typical qubit frequencies.

\section{Model and simulations}

The strength of the damping in Josephson circuits is determined by
power losses at the frequency of plasma resonance $\omega_p = (L_J
C_J)^{-1/2}$, where $L_J = \Phi_0/(2\pi I_c)$ is the Josephson
inductance, $\Phi_0$ the flux quantum, $I_c$ the critical current
and $C_J$ the junction capacitance. When the junction is shunted by
an ohmic resistor (Fig.\,\ref{fig:all-shunts}(a)), the damping is
characterized by the dimensionless McCumber-Stewart parameter
\cite{McCum,Stew},
\begin{equation} \label{beta_c}
\beta_c = (R/\omega_p L_J)^{2} = (2\pi /\Phi_0)I_c R^2 C_J.
\end{equation}
To ensure a normal operation of an RSFQ circuit, its value must be
sufficiently small, i.e. $\beta_c\lesssim 2$ (overdamped regime). In
the case of frequency-dependent shunting (shown in
Fig.\,\ref{fig:all-shunts}(b)-(e)), the resulting damping depends on
the frequency dispersion of the shunting element and on the
parameters of the junction, i.e. $L_J$ and $C_J$. Generally, the
complex admittance of the shunt,
 $Y(\omega)=Y'(\omega)+jY''(\omega)$, adds not only the losses $Y'(\omega)$,
 but causes also a shift of the plasma resonance frequency $\omega_p \rightarrow
 \tilde{\omega}_p$ (see, e.g., Ref.\,\cite{Z-PRB}).
 This effective plasma frequency $\tilde{\omega}_p$ is
 the root of the following equation:
\begin{equation} \label{equation_wp}
\omega [C_J + C^{*}(\omega)] -1/(\omega L_J)= 0,
\end{equation}
where $C^{*}(\omega) = Y''(\omega)/\omega$. The effective
dimensionless damping parameter is, therefore, equal to
\begin{equation} \label{beta_c-eff}
\tilde{\beta}_c =(R^{*}/\tilde{\omega}_p L_J)^{2} =
[\tilde{\omega}_p L_J Y'(\tilde{\omega}_p)]^{-2},
\end{equation}
and this value may reduce to the desirable level of about 1 when the
effective shunting resistance $R^{*}(\omega) \equiv 1/Y'(\omega)$
approaches the high-frequency values $\omega \sim \tilde{\omega}_p$
which lie above the roll-off frequency $\sim(RC)^{-1}$.

\begin{figure}[t]
\begin{center}
\includegraphics[width=8cm]{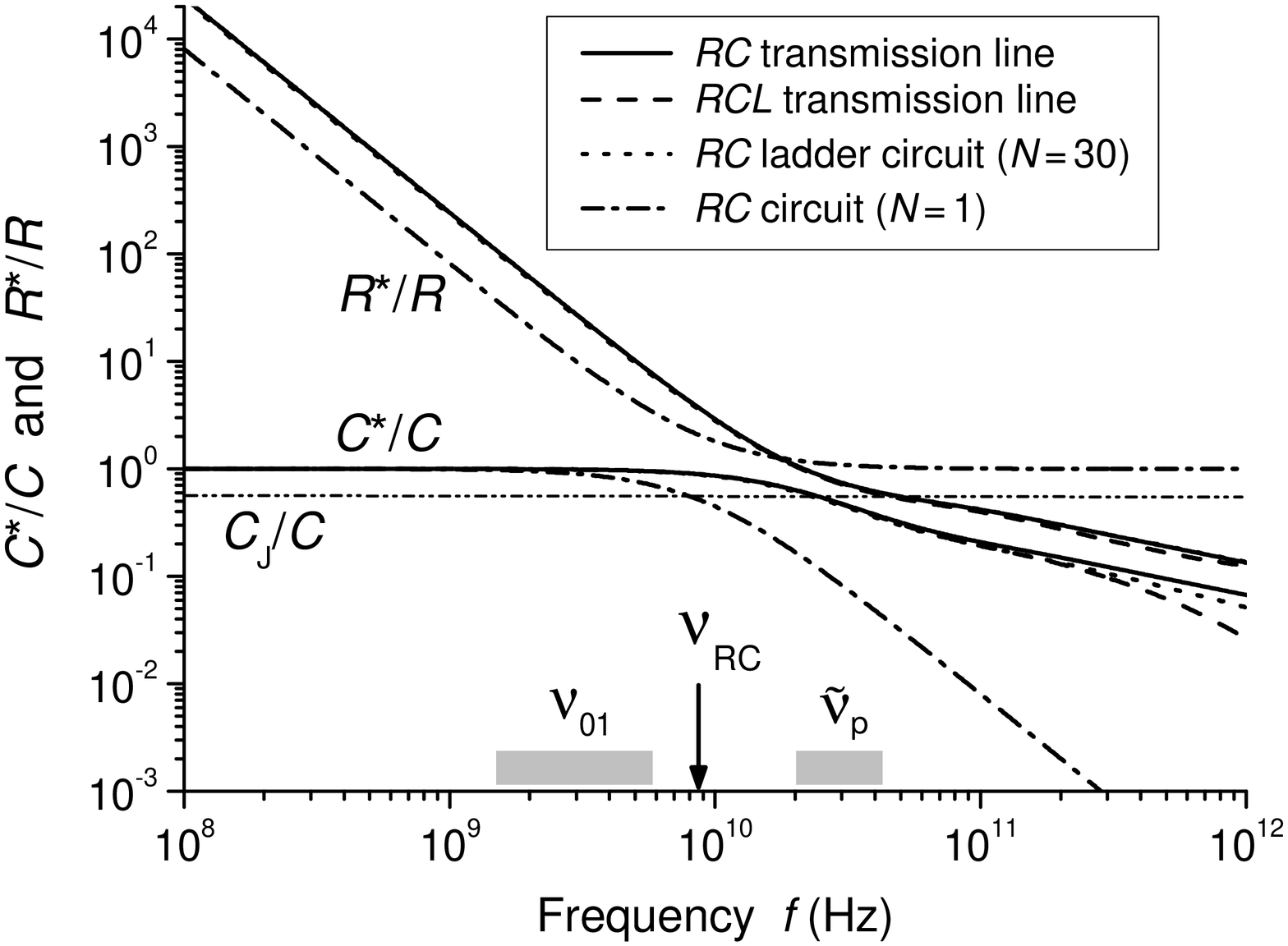}
\caption{Frequency dependencies of the normalized effective
capacitance $C^{*}$ and the normalized effective resistance $R^{*}$
in different types of
shunting circuits, calculated for $R = 10\,\Omega$ and $C =
1.8$\,pF, yielding the transition frequency $\nu_{RC}\equiv (2\pi
RC)^{-1}\approx 8.8$\,GHz. The inductance of the distributed $RCL$
transmission line (dashed lines) is $L=1.9$\,pH. For comparison,
dashed-dotted lines show the behavior of $C^{*}$ and $R^{*}$ in a
simple $RC$ circuit with similar $R$ and $C$. The thin
dashed-dotted-dotted horizontal line shows the value of the junction
capacitance $C_J$. The grey color bars show the typical frequency
ranges of a qubit $\nu_{01}$ and of an RSFQ circuit
$\tilde{\nu}_p$.} \label{fig:RCimp}
\end{center}
\end{figure}

The open-ended transmission lines ($Y_{\textrm{load}}=0$) with the
total resistance $R$, the total capacitance $C$ and the total
inductance $L$, which we apply as Josephson junction shunts, have
the following parameters per unit length: $r=R/\ell$, $c=C/\ell$ and
$l=L/\ell$. The length $\ell$ is smaller than the wavelength
$\lambda_p$ corresponding to the characteristic frequencies ($\sim \tilde{\omega}_p$)
of the problem. We exclude from our analysis the leakage of dielectric and
will normally omit the effect of the line inductance $L$ which, as
we show below, manifests itself at frequencies well above the
operation frequency of the RSFQ circuit. Applying the
transmission-line theory, the admittance seen by the junction is
expressed as
\begin{equation} \label{Y-vs-w}
Y(\omega) = (j\omega C/R)^{1/2} \tanh[(j\omega RC)^{1/2}].
\end{equation}
The plots of the normalized effective shunting resistance
$R^{*}(\omega)/R$ and the effective capacitance $C^{*}(\omega)/C$
calculated using Eq.\,(\ref{Y-vs-w}) are presented in
Fig.\,\ref{fig:RCimp} by solid lines.

The parameters of the frequency-dependent shunting circuits, whose
characteristics are shown in Fig.\,\ref{fig:RCimp}, were chosen such
that the characteristic frequency of the qubit $\nu_{01}$ (of
several GHz) lay notably below the effective plasma frequency
$\tilde{\nu}_p = \tilde{\omega}_p/2\pi \approx 20-40$\,GHz. This
value of $\tilde{\nu}_p$ was estimated from the designed value of
the critical current $I_c= 12\div 24\,\mu$A, the self-capacitance of
the junctions $C_J = 0.5\div 1$\,pF (corresponding to the junction
area $A=12\div 24\,\mu$m$^2$ and the critical current density $j_c =
100$\,A/cm$^2$ in the Nb multilayer process of PTB
\cite{PTBprocess}). The transmission line was realized in the
resistive composite layer consisting of a sandwich Cr/Pt/Cr.
The nominal sheet resistance of this
film was $R_{\Box} = 2\,\Omega$.

The numerical evaluation of $R$, $C$ and $L$ in the transmission
line with the lateral dimensions 30\,$\mu$m by 150\,$\mu$m yielded
an optimum resistance value of $10\,\Omega$ and an optimum
capacitance of $1.8$\,pF. Evaluation of the transmission line
inductance applying the Maxwell software package \cite{Boyko}
yielded the sufficiently small value of $L=1.9$\,pH. The effect of
this inductance, as can be seen from Fig.\,\ref{fig:RCimp} (the
corresponding curves are shown by dashed lines), is negligibly small
in the interesting range of frequencies (up to 200-300\,GHz,
exceeding the plasma frequency $\tilde{\nu}_p$ by one order of
magnitude). Therefore, in our circuit simulations, we assumed a zero
inductance $L=0$ which resulted in a pure diffusive behavior of the
transmission lines. A ladder network model shown in
Fig.\,\ref{fig:all-shunts}(e) with the number of sections $N=30$
gave very good approximation of the distributed line (dotted lines
in Fig.\,\ref{fig:RCimp}). This model made it possible to easily
include these lumped-element networks into the software package
PSCAN \cite{PSCAN} and to simulate the behavior of our circuits with
frequency-dependent damping. For the sake of a further
simplification of the modeling, the circuit with $N=10$ sections
(having very similar characteristics in the interesting range of
frequencies, not shown in Fig.\,\ref{fig:RCimp}) was also
extensively applied in the circuit optimization.

\begin{figure}[t]
\begin{center}
\includegraphics[width = 7.5cm]{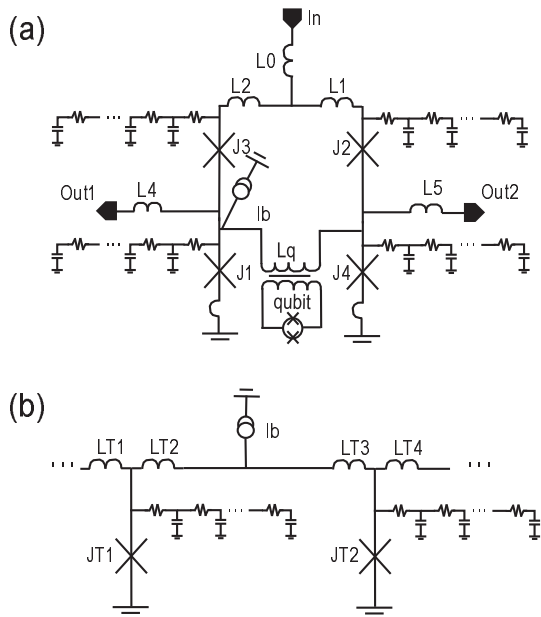}
\caption{Schematics of the basic RSFQ circuits with $RC$
transmission line shunts: (a) TFF, intended for the control of a
Josephson qubit in loop configuration, with a possible inductive
coupling to it (shown in the middle) and (b) one section of JTL.}
\label{TFF-JTL-RCshunt}
\end{center}
\end{figure}

\begin{figure}[t]
\begin{center}
\includegraphics[width = 8cm]{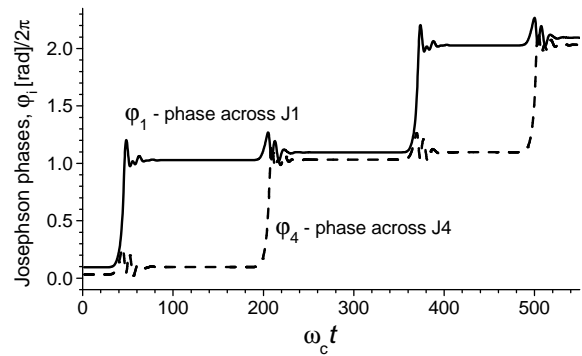}
\caption{Josephson phases across the junctions J1 and J2 of TFF with
$RC$ line shunts (Fig.\,\ref{TFF-JTL-RCshunt}(a)), obtained by
numerical simulations. The characteristic circular frequency used
for normalizing the time scale is $\omega_c =  (20
\,\textrm{ps})^{-1}$.} \label{TFFsim}
\end{center}
\end{figure}

We have modeled the behavior of the basic RSFQ circuits including
TFF (shown in Fig.\,\ref{TFF-JTL-RCshunt}(a)) and JTL
(see Fig.\,\ref{TFF-JTL-RCshunt}(b)) with $RC$ transmission line shunts.
Such TFF circuit has been developed for the control of the Josephson
qubit in SQUID configuration so its quantizing inductance $L_q$ is
magnetically coupled to the qubit loop. As has recently been
demonstrated in Ref.\,\cite{Castellano}, the output rectangular
magnetic flux pulses of TFF, triggered by the incoming SFQ pulses at
the port "In", can effectively change the shape of the energy
potential of the double-SQUID flux qubit and, hence, form the basis
for the quantum manipulation. For this type of Josephson qubit, a
more than 300-fold increase in the effective shunting resistance can
be achieved at its typical frequency $\nu_{01} \approx 1$\,GHz. The
absolute value of $R^{*}$ in our design is 3\,k$\Omega$ at this frequency.
Note that according to estimations made by Chiarello \cite{Chiarello} taking
into account the effect of attenuation of noise due to sufficiently
weak coupling, the resistance $R^{*}$ kept at sufficiently low
temperature of 20\,mK must not be lower than 100\,$\Omega$. In this
case, decoherence of the qubit is still tolerable. Therefore, the
designed $RC$ transmission line shunts are very well-suited for
controlling this qubit.

The result of the simulations for the set of parameters typical for
our technology is shown in Fig.\,\ref{TFFsim} as time dependencies
of the Josephson phases reacting on the input SFQ pulses. The
step-like behavior of the phases $\varphi_1$ and $\varphi_4$ on the
Josephson junctions $J$1 and $J$4, respectively, is due to $2\pi$
phase leaps corresponding to the switching of the TFF between two
states. An incoming SFQ pulse (not shown here) induces 2$\pi$ phase
leaps across the junctions $J$1 and $J$2 (see
Fig.\,\ref{TFF-JTL-RCshunt}); the next SFQ pulse causes 2$\pi$ leaps
across junctions $J$3 and $J$4. This yields the sequence of current
pulses in the storing inductance $L_q$, and therefore the pulses
of the output magnetic flux. These pulses have required rectangular shape
(see, for comparison, similar plots in Fig.\,7(b) of
Ref.\,\cite{Z-PRB}, obtained for SIN-junction-shunted TFF). The
parameter margins found in the optimization of the circuit are in
the safe range typical of RSFQ circuits \cite{Intiso}; the most
critical parameter is the critical current density $j_c$, having
margins of $\pm 28 \%$.

\section{Experiment}

\begin{figure}[b]
\begin{center}
\includegraphics[width = 7.5cm] {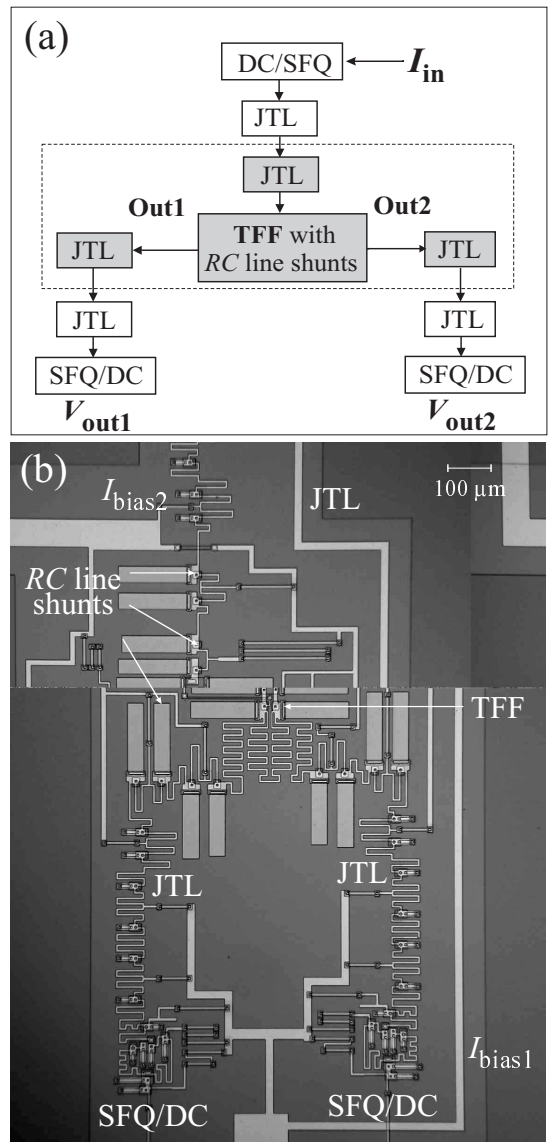}
\caption{(a) Block diagram of the circuit with $RC$ line shunts
(elements inside the dashed-line box) and resistive shunts (outside
this box) including TFF, JTL, DC/SFQ and SFQ/DC converters. (b)
Image of the circuit fabricated at PTB in Nb technology with $j_c
=100$\,A/cm$^2$. Critical currents of the TFF junctions are $I_c =
17\,\mu$A each, the quantizing inductance $L_q = 132$\,pH.}
\label{TFF-diag-image}
\end{center}
\end{figure}

The circuit designed for the experimental testing of the
functionality of the $RC$ transmission line shunts is the TFF cell
which is similar to that shown in Fig.\,\ref{TFF-JTL-RCshunt}, but
its input and two outputs were attached to JTLs with $RC$ line
shunts serving as buffer stages. In the block diagram in
Fig.\,\ref{TFF-diag-image}(a), this part of the circuit is marked by
grey color and placed inside the dashed-line box. Due to the
frequency-dependent damping, such a JTL can transmit SFQ pulse
signals and efficiently attenuate noise originating from the
external part of the circuit. Each JTL circuit includes two sections
which ensure an attenuation of the noise coming from the
conventional resistive shunts of an external circuit by more than
40\,dB. Such a large attenuation was possible due to the relatively
low currents biasing the sections of the JTL, which resulted in rather
small values of the Josephson inductances and, therefore, a division
of signals with large ratio. The "external" part of the circuit
includes four-section JTLs with resistive shunts as well as DC/SFQ
and SFQ/DC converters. The DC/SFQ converter is used for generating
the sequence of the SFQ pulses driving the TFF, while two SFQ/DC
converters enable visualization of the TFF switching. The SFQ pulses
transmitted via two output channels of the TFF are therefore
converted into rectangular voltage pulses. The edges of these
rectangular pulses are time-referenced to the input SFQ pulses and
the repetition frequency is four times lower than the frequency of
the input pulses. Moreover, the two output signals have a mutual
phase delay. This behavior must originate from the pulse rate
division by two sequentially connected TFFs. The first one is the
core TFF with $RC$ transmission-line-shunted junctions, and the
second one in each line is the TFF of conventional design with
integrated SFQ pulse detector with resistively shunted junctions.
The mutual shift of the output signals $V_{\textrm{out}1}$ and
$V_{\textrm{out}2}$ should correspond to exactly one period of the
input signal $I_{\textrm{in}}$.

The fabrication process included the deposition of an Nb film (170
nm thick) which was patterned for structuring a ground plane of the
circuits. The electrical insulation of the ground plane was provided
by a wet anodization of the Nb film, followed by the sputtering of a
dielectric SiO$_{2}$ layer (80 nm thick) on top of it. To form the
bias and shunt resistors, we deposit a Cr/Pt/Cr (10 nm/43 nm/10 nm)
sandwich patterned by an Ar-milling etch process. Thereafter, a
second SiO$_{2}$ ground plane insulation layer (140 nm thick) was
sputtered. In order to connect the groundplane and the resistors to
other metal layers, holes were etched in this dielectric layer. The
Nb/Al-Al$_{x}$O$_{y}$/Nb (170 nm/10 nm/80 nm) trilayer was deposited
in an UHV system with a base pressure lower than $10^{-6}$\,Pa. The
oxidation process was performed in a load-lock with a base pressure
lower than $10^{-5}$\,Pa. The tunnel areas of the Josephson junctions
were formed by etching the Nb counter electrode of the trilayer.
Then, the wet anodization process was applied again, followed by the
patterning of the Nb base electrode. An SiO$_{2}$ (300 nm) layer was
sputtered onto the sample for insolating the edges of the base
electrode and strengthening the anodic oxide. The process was
completed by sputtering an Nb (400 nm) wiring layer. The optical
microscope image of a fragment of the finished sample is shown in
Fig.\,\ref{TFF-diag-image}(b). The $RC$ transmission line shunts of
rectangular shape are seen in this image.

\begin{figure}[t]
\begin{center}
\includegraphics[width = 8cm] {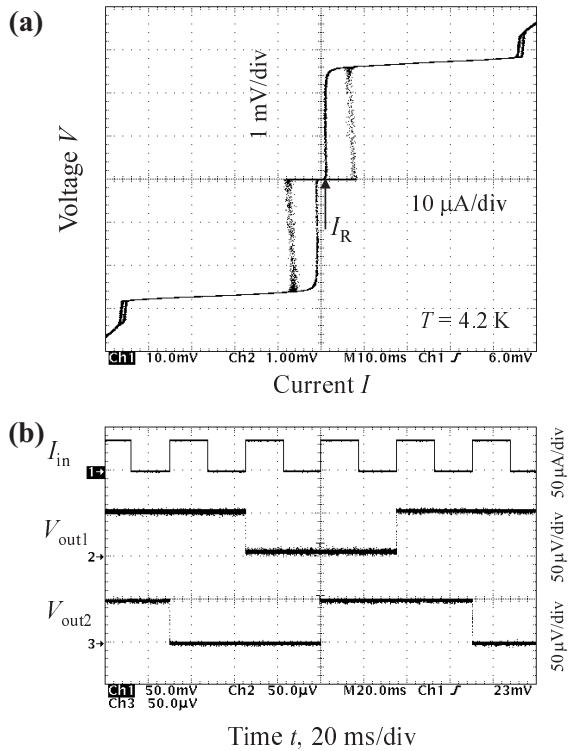}
\caption{(a) Autonomous $I$-$V$ curve of the $RC$-line-shunted
Josephson junction. (b) Time traces of the signals in the TFF
circuit with $RC$-line-shunted junctions. The frequency of the
output signals $V_{\textrm{out1}}$ and $V_{\textrm{out2}}$ is four
times lower compared with the frequency of the input signal
$I_{\textrm{in}}$. The observed division of the frequency of the
input pulses indicates that the TFF operates correctly.}
\label{TFF-exp-curves}
\end{center}
\end{figure}

The measurements of the samples were performed at a liquid helium
temperature of $T=4.2$\,K. At first, we tested the stand-alone
Josephson junctions with $RC$ transmission line shunts which were
fabricated on the same chip for the purpose of measuring their
autonomous $I$-$V$ characteristics. Figure\,\ref{TFF-exp-curves}(a)
shows a typical $I$-$V$ curve, exhibiting rather strong hysteresis.
The critical current $I_c$ with a nominal value of 16.6$\,\mu$A is
notably suppressed due to the noise in the measuring lines, so the
switching into the resistive state occurs at $I \approx 8\,\mu$A.
The value of the return current $I_R$, determined by both the
junction capacitance $C_J$ and the shunting admittance $Y(\omega)$,
is about $1\,\mu$A. Such a value of $I_R$ corresponds in the
resistively-shunted model of the junction \cite{McCum,Stew} to the
value of $\beta_c \approx (4I_c/\pi I_R)^2 \approx 200$. In the
quiescent (zero-voltage) state of the circuit, the zero-bias
quasiparticle resistance $R_{\textrm{qp}}$ has to be taken into
account for evaluating the low-frequency noise. The values of this
resistance are estimated to be higher than 2\,k$\Omega$ for the
stand-alone junctions and 1\,k$\Omega$ for the junctions used in the
TFF circuit. Note that the equivalent value of $\beta_c$, derived
for an unshunted junction with $R_{\textrm{qp}}=1$\,k$\Omega$, is
$5\times 10^4$, while the effective value $\tilde{\beta}_c$ designed
for the junction shunted by the $RC$ transmission line is about 1.
This value ensures a sufficient damping in the dynamic regime.

Figure\,\ref{TFF-exp-curves}(b) demonstrates the signal time traces
in the circuit shown in Fig.\,\ref{TFF-diag-image}. The output
signals $V_{\textrm{out1}}$ and $V_{\textrm{out2}}$ shifted with
half a period and with frequency four times lower than the frequency
of the input signal $I_{\textrm{in}}$ confirm that the circuits
operate correctly. The range of its operation was found to be
sufficiently wide. The margins for the bias currents were $\pm 21\%$
for the $RC$-line-shunted TFF circuit ($I_{\textrm{bias2}}$, see
Fig.\,\ref{TFF-diag-image}(b)) and $\pm 30\%$ for the bias of the DC/SFQ and
SFQ/DC converters based on resistively shunted junctions
($I_{\textrm{bias1}}$).

\section{Discussion}

One of the complex issues in designing an integrated "RSFQ + qubit"
circuit operating at millikelvin temperature is the overheating of
the shunting resistors due to the dissipated power $P\approx f I_c
\Phi_0$, where $f$ is the working (clock) frequency \cite{Savin2}. Because
of the rather small volume $\Lambda_R \approx 10\,\mu$m$^{3}$ of
conventional thin film resistors (they must have a relatively small
stray capacitance) the effective electron temperature $T_e\approx
(P/\Sigma \Lambda_R)^{1/5}$ (here $\Sigma \sim
1$\,nW\,$\mu$m$^{-3}\,$K$^{-5}$ is the material constant) may
significantly exceed the bath temperature even at low $f$
\cite{Savin}. The efficient method proposed in Ref.\,\cite{Savin} of
reducing $T_e$ was based on the application of bulky metallic
reservoirs ($\Lambda_{f} \approx 4\times 10^{5}\,\mu$m$^{3}$),
so-called cooling fins, attached to the small resistors. The volume
of the $RC$ line shunts designed in our paper is $\Lambda_{RC}
\approx 284\,\mu$m$^{3}\approx 30 \Lambda_R$, and this increased
volume should result in an approximately two-fold
($30^{\frac{1}{5}}\approx 2$) reduction in their electron
temperature $T_e$ without attaching them to cooling fins. Moreover,
the large size of the $RC$ line shunts makes it, in principle,
possible to attach cooling fins of larger volume. These cooling fins
may have a more efficient thermal contact to the large "hot" area of
the $RC$ line shunt than in the case of relatively small resistive
shunts or the $RC$ shunts \cite{VTT-2}.

Another advantage of the $RC$ transmission line shunts over the $RC$
shunts consists in inherently diffusive transmission characteristics
of these lines and, therefore, the absence of parasitic
electromagnetic resonances due to standing waves in these
structures. These resonances may present a problem in Nb-based
on-chip capacitors having very low losses. (The $RC$ shunts with
such capacitors were realized in VTT technology with a critical
current density $j_c = 30$\,A/cm$^{2}$ \cite{VTT-2}.) Therefore, the
lateral dimensions of the capacitors (100\,$\mu$m by 160\,$\mu$m
with 140\,nm thick Nb$_{2}$O$_{5}$ dielectric yielding $C=35$\,pF
\cite{VTT-2}) should not exceed $\lambda_p/8$ ($\approx 170\,\mu$m),
where $\lambda_p$ is the electromagnetic wavelength in the
dielectric at the plasma frequency.

Comparison of both the levels of the low-frequency noise and the
efficiencies of the high-frequency damping in the $RC$ transmission
line shunt and $RC$ shunt yields the following results. For similar
values of the parameters $R$ and $C$ in both models, the effective
conductance $Y' = 1/R^{*}$ at $\omega \ll (RC)^{-1}$ is equal to
$\omega^{2}C^{2}R$ and $\omega^{2}C^{2}R/3$ (given by
Eq.\,(\ref{Y-vs-w}) in the low-frequency limit), respectively. This
difference is also to be seen in the plot in Fig.\,\ref{fig:RCimp}.
For the equal effective temperatures of these shunts, this yields a
three-times lower spectral density of the current noise ($=
4k_BT_eY'$). There is also a difference in the behavior at high
frequencies, $\omega \gg (RC)^{-1}$. In this frequency range, the
effective capacitance of the system "junction + shunt" is mostly
determined by the capacitance of the junction $C_J$ (compare the
behavior of $C^{*}$ shown by the solid and dashed-dotted lines with
the typical level of $C_J$ shown by thin dashed-dotted-dotted line
in Fig.\,\ref{fig:RCimp}). Therefore, the strength of the damping is
determined by the shunt losses only. The $RC$ transmission line
model yields, therefore, the appreciably larger damping, $R^{*}=
(2R/\omega C)^{1/2} \ll R$ (high-frequency limit in
Eq.\,(\ref{Y-vs-w})) than the plain damping $R$ in the $RC$ shunt
model. The enhanced high-frequency damping should stabilize the
operation of the RSFQ circuits with the $RC$ transmission line
shunts.

In conclusion, we developed an efficient approach to the problem of
the frequency-dependent damping of RSFQ circuits to be used for
integration with Josephson qubits. The proposed junction shunts
based on $RC$ transmission lines have remarkable electric
characteristics and, as we have demonstrated, can be easily
implemented in the framework of the available Nb technology
\cite{PTBprocess}.

\section{Acknowledgements}

We are grateful to Boyko Dimov for assistance in simulations and
Friedrich-Immanuel Buchholz, Gabriella Castellano and Fabio Chiarello
for helpful discussions. This work was
partially supported by the EU through the RSFQubit and EuroSQIP
projects.


\end{document}